\documentclass[aps,prb,preprint,superscriptaddress]{revtex4}
\usepackage{amssymb}
\usepackage{graphicx}
\usepackage{amsmath}
\usepackage{color}

\begin{document}
\title{Spin-orbit driven ferromagnetic resonance: A nanoscale magnetic characterisation technique}

\author{D.~Fang}
\affiliation{Microelectronics Group, Cavendish Laboratory, University of Cambridge, JJ Thomson Avenue, Cambridge CB3 0HE, UK}
\author{H.~Kurebayashi}
\affiliation{Microelectronics Group, Cavendish Laboratory, University of Cambridge, JJ Thomson Avenue, Cambridge CB3 0HE, UK}
\author{J.~Wunderlich}
\affiliation{Hitachi Cambridge Laboratory, Cambridge CB3 0HE, UK}
\affiliation{Institute of Physics ASCR, v.v.i., Cukrovarnick\'{a} 10, 162 53 Praha 6, Czech Republic}
\author{K.~V\'yborn\'y}
\affiliation{Institute of Physics ASCR, v.v.i., Cukrovarnick\'{a} 10, 162 53 Praha 6, Czech Republic}
\author{L.~P.~Z${\rm \hat{a}}$rbo}
\affiliation{Institute of Physics ASCR, v.v.i., Cukrovarnick\'{a} 10, 162 53 Praha 6, Czech Republic}
\author{R.~P.~Campion}
\affiliation{School of Physics and Astronomy, University of Nottingham, Nottingham NG7 2RD, UK}
\author{A.~Casiraghi}
\affiliation{School of Physics and Astronomy, University of Nottingham, Nottingham NG7 2RD, UK}
\author{B.~L.~Gallagher}
\affiliation{School of Physics and Astronomy, University of Nottingham, Nottingham NG7 2RD, UK}
\author{T.~Jungwirth}
\affiliation{Institute of Physics ASCR, v.v.i., Cukrovarnick\'{a} 10, 162 53 Praha 6, Czech Republic}
\affiliation{School of Physics and Astronomy, University of Nottingham, Nottingham NG7 2RD, UK}
\author{A.~J.~Ferguson}
\email{ajf1006@cam.ac.uk}
\affiliation{Microelectronics Group, Cavendish Laboratory, University of Cambridge, JJ Thomson Avenue, Cambridge CB3 0HE, UK}

\begin{abstract}
We demonstrate a scalable new ferromagnetic resonance (FMR) technique based on the spin-orbit interaction. An alternating current drives FMR in uniform ferromagnetic structures patterned from the dilute magnetic semiconductors (Ga,Mn)As and (Ga,Mn)(As,P). This allows the direct measurement of magnetic anisotropy coefficients and damping parameters for individual nano-bars. By analysing the ferromagnetic resonance lineshape, we perform vector magnetometry on the current-induced driving field, observing contributions with symmetries of both the Dresselhaus and Rashba spin-orbit interactions.
\end{abstract}

\maketitle

Ferromagnetic resonance (FMR) is the most common technique for exploring spin-dynamics phenomena and for the magnetic characterisation of ferromagnets.\cite{vonsovskii} However, previously developed FMR techniques, based on exciting the magnetic system by an external alternating magnetic field from a resonant cavity\cite{goennenwein:apl07,mecking:prb07,hui:apl08} or a micro-waveguide,\cite{costache:apl06,costache:apl06-2,yamaguchi:prb08,hoffmann:prb09} struggle to simultaneously achieve scalability of the technique to nano-size objects, uniformity of the excitation field, and the range of available excitation frequencies. We introduce an FMR technique applicable to individual nanomagnets in which the FMR driving field is generated in the probed magnet itself. The excitation is driven by the effective field generated by an alternating electrical current passing through the ferromagnet, which results from the combined effect of the spin-orbit (SO) coupling and exchange interaction. \cite{Manchon:prb09,rokhinson:naturep09,miron:natmat10}  Our SO-FMR can be operated at tuneable frequencies and we demonstrate its sensitivity and scalability by measuring the variation of micromagnetic parameters of lithographically patterned (Ga,Mn)As and (Ga,Mn)(As,P) nano-bars.

FMR induced by driving an alternating current directly through the probed sample has been previously demonstrated for specific non-uniform magnetic nano-devices such as spin-valves.\cite{Tulapurkar:nature05,sankey:prl06} The experiments utilised the spin-transfer torque in which spin-polarised electrical current acts on spatially varying magnetisation\cite{{myers:science99}} and can be viewed as a macroscopic angular momentum transfer effect. Our SO-FMR (Figure~1a) does not require the specific samples with a non-collinear magnetisation profile. The method can be applied to a broad range of systems including uniformly polarised nanomagnets. This is because the effective field utilised in the SO-FMR does not rely on the spatial variation of the magnetisation vector but on a microscopic non-collinearity of individual electron spins due to their relativistic, SO-coupled band structure. Specifically, when an electrical current traverses through the uniformly magnetised material, the resulting non-equilibrium distribution of occupied states in the SO-coupled carrier bands yields a non-equilibrium spin polarisation.\cite{edelstein,inoue:prb03,silov:apl04} The polarisation produces a transverse component of the internal exchange field (can be viewed as an effective magnetic field) and a torque is applied to the magnetisation vector.\cite{Manchon:prb09,garate:prb09} This current-induced effective field is generic to ferromagnets with SO-coupling and inversion asymmetry in their band structure. Previously it has been utilised for magnetisation switching in the ferromagnetic semiconductor (Ga,Mn)As\cite{rokhinson:naturep09} and for domain nucleation in a Pt/Co/AlO$_x$ stack.\cite{miron:natmat10}

The micro and nano-bars employed in our SO-FMR study are patterned by electron beam lithography on 25~nm-thick films of (Ga$_{0.94}$,Mn$_{0.06}$)As and (Ga$_{0.94}$,Mn$_{0.06}$)(As$_{0.9}$,P$_{0.1}$), grown by low-temperature molecular beam epitaxy. The (III,Mn)V ferromagnetic semiconductors used in our study are particularly favourable systems for observing and exploring SO-FMR because of the compatibility of the material with advanced semiconductor nanofabrication techniques, because the carrier bands have strong SO-coupling, and the (III,Mn)V nanostructures have a rich phenomenology in their micromagnetic parameters. In the following text we demonstrate our scalable SO-FMR technique in lithographically patterned bars of width ranging from several $\mu$m's to 80~nm (Figure~1b).

In order to drive SO-FMR we pass a microwave-frequency current through the nano-bar. This is achieved by wire-bonding the sample between an open-circuit coplanar transmission line and a low-frequency connection which also provides a microwave ground (Figure~1c). Since the microwave excitation field originates from the material properties, only a 2-terminal nano-bar (a resistor) is required in our experiment, enabling simple and rapid sample fabrication. For detection of FMR we utilise a frequency mixing effect based on the anisotropic magnetoresistance (AMR).\cite{costache:apl06,costache:apl06-2,goennenwein:apl07,mecking:prb07,hui:apl08,yamaguchi:prb08} When the magnetisation precession is driven, a time-dependent change $\Delta R(t)$ in the longitudinal resistance from the equilibrium value $R$ occurs (due to the AMR). The resistance oscillates with the same frequency as the microwave current, therefore causing frequency mixing and a directly measurable dc voltage $V_\text{dc}$ is generated across the nano-bar. This voltage is our observable providing a direct probe of the amplitude and phase of the magnetisation precession with respect to the microwave current.

We first show measurements on a 80~nm-wide nano-bar patterned in the [1\={1}0] direction from the (Ga,Mn)(As,P) epilayer. The magnetic field dependence of $V_\text{dc}$ is measured at different microwave frequencies and taken at a temperature of 6~K. The frequency of the incident current is fixed while an external dc magnetic field $\mathbf{H}_0$ is swept and a well-defined resonance peak appears (Figure~2a). The peak is well-fitted by the solution of the Landau-Lifshitz-Gilbert (LLG) equation, which describes the dynamics of precessional motion of the magnetisation:
    \begin{equation}
        V_\text{dc} = V_\text{sym}\frac{\Delta H^2}{(H_0 - H_\text{res})^2 + \Delta H^2} + V_\text{asy}\frac{\Delta H(H_0 - \Delta H)}{(H_0 - H_\text{res})^2 + \Delta H^2}
            \label{eq:Vdc}
    \end{equation}
Here $H_\text{res}$ is the field at which resonance occurs and $\Delta H$ is the linewidth (half width at half maximum) of the FMR peak. The resonance lineshape is a combination of symmetric and anti-symmetric Lorentzian functions with amplitudes $V_\text{sym}$ and $V_\text{asy}$, respectively. Their relative contributions are determined by the phase of the driving field with respect to the current, and the direction of the driving field (see Equation~\ref{eq:Vsym} \& \ref{eq:Vasy}).

Figure~2b plots the frequency-dependence of the resonance field $H_\text{res}$. It is described by the equation for ferromagnetic resonance:\cite{liu:condmat06}
 \begin{equation}
   \left( \frac{\omega}{\gamma}\right)^2 =  \mu_0^2(H_\text{res} + H_\text{ani}^{'})(H_\text{res}+H_\text{ani}^{''})
        \label{eq:resonance}
 \end{equation}
where $H_\text{ani}^{'}$ and $H_\text{ani}^{''}$ are terms containing the demagnetisation and anisotropy energies of the ferromagnet (see Methods). A gyromagnetic constant $\gamma$ characteristic for Mn$^{2+}$ spins of 176~GHz/T (g-factor 2) is used for the fitting. This, together with the good agreement between the observed peaks and the fitted results from the LLG equation, confirms that we observe the coherent precession of Mn spins.

The FMR linewidth ($\Delta H = \Delta H_\text{inhomo} + \alpha\omega/\gamma$) describes the damping in the ferromagnetic system. The broadband nature of our setup allows us to determine the inhomogeneous (2.5~mT) and frequency-dependent contributions to the damping (Figure 2c) that correspond to Gilbert-damping constant $\alpha=$~0.023.  Using a vector field cryostat we also perform the SO-FMR measurements for different orientations of the external magnetic field. In Figure~2d we present the data from an in-plane scan of the magnetic field showing that there is a strong uniaxial anisotropy perpendicular to the bar direction. By analysing the peak positions (Figure~2e) using Equation~\ref{eq:resonance} we quantify the anisotropy fields and find $\mu_0H_{2\parallel}=-180$~mT (uniaxial) and $\mu_0H_{4\parallel}=68$~mT (biaxial).

We now demonstrate that SO-FMR can be applied to comparative investigations of nano-bars where the anisotropies differ from bulk values. The effect of strain-relaxation, due to the lithographic patterning, on the magnetic anisotropy of (Ga,Mn)As nano-bars has previously been studied by electrical transport\cite{humpfner:apl07,wunderlich:prb07,wenisch:prl07} and optically-detected FMR.\cite{hoffmann:prb09} We first compare the effect of strain-relaxation between 500~nm bars under compressive ((Ga,Mn)As) and tensile ((Ga,Mn)(As,P)) growth strain. The in-plane anisotropies are studied; although (Ga,Mn)(As,P) is out-of-plane magnetised\cite{rushforth:jap08}, the applied field $\mathbf{H}_0$ brings the magnetisation into plane. In (Ga,Mn)As we observe an additional uniaxial contribution to the anisotropy ($\mu_0H_U=32$~mT) along the bar (Figure 3a \& c) with a similar magnitude to previous reports.\cite{humpfner:apl07,wenisch:prl07,hoffmann:prb09} By contrast in the (Ga,Mn)(As,P) nano-bar (Figure 3b \& c) the sign of the uniaxial anisotropy ($\mu_0H_U=-50.1$~mT) has reversed and the easy axis is now perpendicular to the bar. This can be understood in terms of the sign of the strain relaxation: these materials become magnetically easier in the direction of most compressive (least tensile) strain. So when the tensile strain of the (Ga,Mn)(As,P) nano-bar relaxes, it introduces an easy axis perpendicular to the bar (Figure 3d). Furthermore we measure (Ga,Mn)(As,P) bars of different widths and observe a decrease in the strain-relaxation induced anisotropy from the 80~nm bar ($\mu_0H_U=-270$~mT) to the 500~nm bar ($\mu_0H_U=-50.1$~mT), and almost no effect of strain-relaxation in the $4~\mu$m bar ($\mu_0H_U=-10.5$~mT).

As well as being able to determine the patterning-induced change in anisotropy, we also compare the damping among the nano-bars of different sizes. The frequency-dependent term (related to damping) increases for decreasing bar width: $\alpha = 0.004$ (4~$\mu$m-wide), 0.006 (500~nm) and 0.023 (80~nm). The significantly higher value of Gilbert damping at 80~nm compared with the 500~nm and 4~$\mu$m bars may be due to damage during the etching process. The frequency-independent term is relevant in the case of strain relaxation as it indicates the inhomogeneity of anisotropy fields within the bar itself. The intermediate case of 500~nm shows greater inhomogeneity $\Delta H_\text{inhomo}=9.9$~mT than the 4~$\mu$m bar $\Delta H_\text{inhomo}=5.4$~mT, explained by the increased variation in local anisotropy. By contrast, for 80~nm bar reduces to $\Delta H_\text{inhomo}=2.5$~mT, indicative of a high degree of strain-relaxation.

To characterise SO-FMR we must understand the direction and amplitude of the effective field $\mathbf{h}_\text{eff}$ that drives magnetisation precession. Similar to the experiments on STT-FMR in spin-valves\cite{Tulapurkar:nature05,sankey:prl06} we are able to perform vector magnetometry on the driving field from the angle dependence of the amplitude of the FMR peak. For a vector driving field $\mathbf{h}_\text{eff}(t) = (h_x, h_y, h_z)e^{i\omega t}$ in-phase with the microwave current $\mathbf{I} (t) = (I_x, 0, 0)e^{i\omega t}$, the amplitudes of the two components of the FMR peak are:
    \begin{eqnarray}
        V_\text{sym} (\theta) &=&  \frac{I \Delta R}{2} A_\text{sym} \sin(2 \theta) h_z     \label{eq:Vsym} \\
        V_\text{asy} (\theta) &=&  \frac{I \Delta R}{2} A_\text{asy} \sin(2 \theta) (h_x\sin\theta + h_y\cos\theta)     \label{eq:Vasy}
    \end{eqnarray}
where $\Delta R$ is the AMR coefficient of the ferromagnetic sample, $\theta$ is the angle between the applied field $\mathbf{H}_0$ and the current $\mathbf{I}$, and $A_\text{sym(asy)}$ are constants determined by the magnetic anisotropies. Hence by decomposing the resonance lineshape into $V_\text{sym}$ and $V_\text{asy}$, and by measurements of the AMR and magnetic anisotropies we are able to deduce the components of $\mathbf{h}_\text{eff}$.

No component of $V_\text{sym}$ is seen to behave as $\sin(2 \theta)$, indicating that the driving field $\mathbf{h}_\text{eff}$ is predominantly in-plane. Figure~4a shows the angle-dependence of $V_\text{asy}$ for a 500~nm-wide (Ga,Mn)As bar patterned in the [1$\bar{1}$0] direction. We see that $V_\text{asy} (\theta)$ comprises a $-\sin(2\theta)\cos(\theta)$ term, indicating that the driving field is perpendicular to \textbf{I}. In a [110] device (Figure~3a) the amplitude of $V_\text{asy}$ has the opposite sign, indicating that the driving field has reversed. For nano-bars along [100] and [010] (Figure~3b), the $V_\text{asy}$ curve is a superposition of $\sin(2\theta)\sin(\theta)$ and $\sin(2\theta)\cos(\theta)$ functions, showing that the driving field consists of components both parallel and perpendicular to $\mathbf{I}$.

These data are most clearly seen by plotting the dependence of the magnitude and direction of the effective field on the current (nano-bar) orientation (Figure~3c). Two contributions to the driving field are observed with different symmetry, $\mathbf{h}_\text{eff} = \mathbf{h}_\text{R} + \mathbf{h}_\text{D} $. The fields $ \mathbf{h}_\text{R}$ and $ \mathbf{h}_\text{D}$ have angular dependence on $\mathbf{I}$ reminiscent of the angular dependence of Rashba and Dresselhaus SO fields in the momentum space, respectively.\cite{dresselhaus,rashba}  The field with Dresselhaus symmetry, as previously observed in magnetisation switching experiments,\cite{rokhinson:naturep09} is due to the diagonal elements in the strain tensor (due to the lattice mismatch between GaAs substrate and (Ga,Mn)As). Therefore $ \mathbf{h}_\text{D}$ changes sign between the (Ga,Mn)As and (Ga,Mn)(As,P) materials (comparing Figure~4c and 4d). The Rashba symmetry field $\mathbf{h}_\text{R}$ can be modelled by off-diagonal elements in the strain tensor. This strain is not physically present in the crystal structure of (Ga,Mn)As epilayers. It has been introduced, however, in previous studies to model the in-plane uniaxial anisotropy present in (Ga,Mn)As and the fitted values of this effective off-diagonal strain are typically several times smaller than the diagonal, growth-induced strain.\cite{zemen:prb09} This is consistent with the observed smaller magnitude of $\mathbf{h}_\text{R} = 6.5$~$\mu$T than $\mathbf{h}_\text{D} = 18$~$\mu$T (values given at $j = 10^5$~Acm$^{-2}$). Both $\mathbf{h}_\text{D}$ and $\mathbf{h}_\text{R}$ are measured to be linear in current density (Figure~4e \& f). We observe a larger magnitude of $\mathbf{h}_\text{D}$ at a given current density in the (Ga,Mn)(As,P) nano-bars. This is explained by the larger magnitude of the growth strain and larger resistivity (larger $E$ at given $j$) of (Ga,Mn)(As,P) as compared to the (Ga,Mn)As film.\cite{rushforth:jap08}

In conclusion, we perform variable-frequency FMR experiments on individual micro and nano-bars of uniform ferromagnetic semiconductors (Ga,Mn)As and (Ga,Mn)(As,P). The FMR is driven by a torque at microwave frequencies whose origin lies in the internal effective field (due to the SO-coupling and exchange interaction) of the probed ferromagnet. We have demonstrated the utility of our SO-FMR technique by determining the rich characteristics of magnetic anisotropy fields and damping coefficients in the studied nanoscale ferromagnetic semiconductor samples. In addition, we have performed vector magnetometry on the driving field allowing us to measure a previously unobserved contribution to the current-induced field in the studied ferromagnets with symmetry of the Rashba SO-interaction. Our work demonstrates a new scalable FMR technique which provides an unprecedented method to perform magnetic characterisation of uniform ferromagnetic nanostructures and to study the nature of the current-induced effective magnetic field in SO-coupled ferromagnets.

We acknowledge fruitful discussions with Ion Garate, Allan H. MacDonald and Leonid Rokhinson and support from EU Grants FP7-214499 NAMASTE, FP7-215368 SemiSpinNet, ERC Advanced Grant, from Czech Republic Grants AV0Z10100521, KAN400100652, LC510, KJB100100802 and Preamium Academiae, DF acknowledges support from Cambridge Overseas Trusts and Hitachi Cambridge Laboratory, A.J.F. acknowledges the support of a Hitachi research fellowship.

\newpage


\newpage

\begin{figure}
	\centering
		\includegraphics[width=10cm]{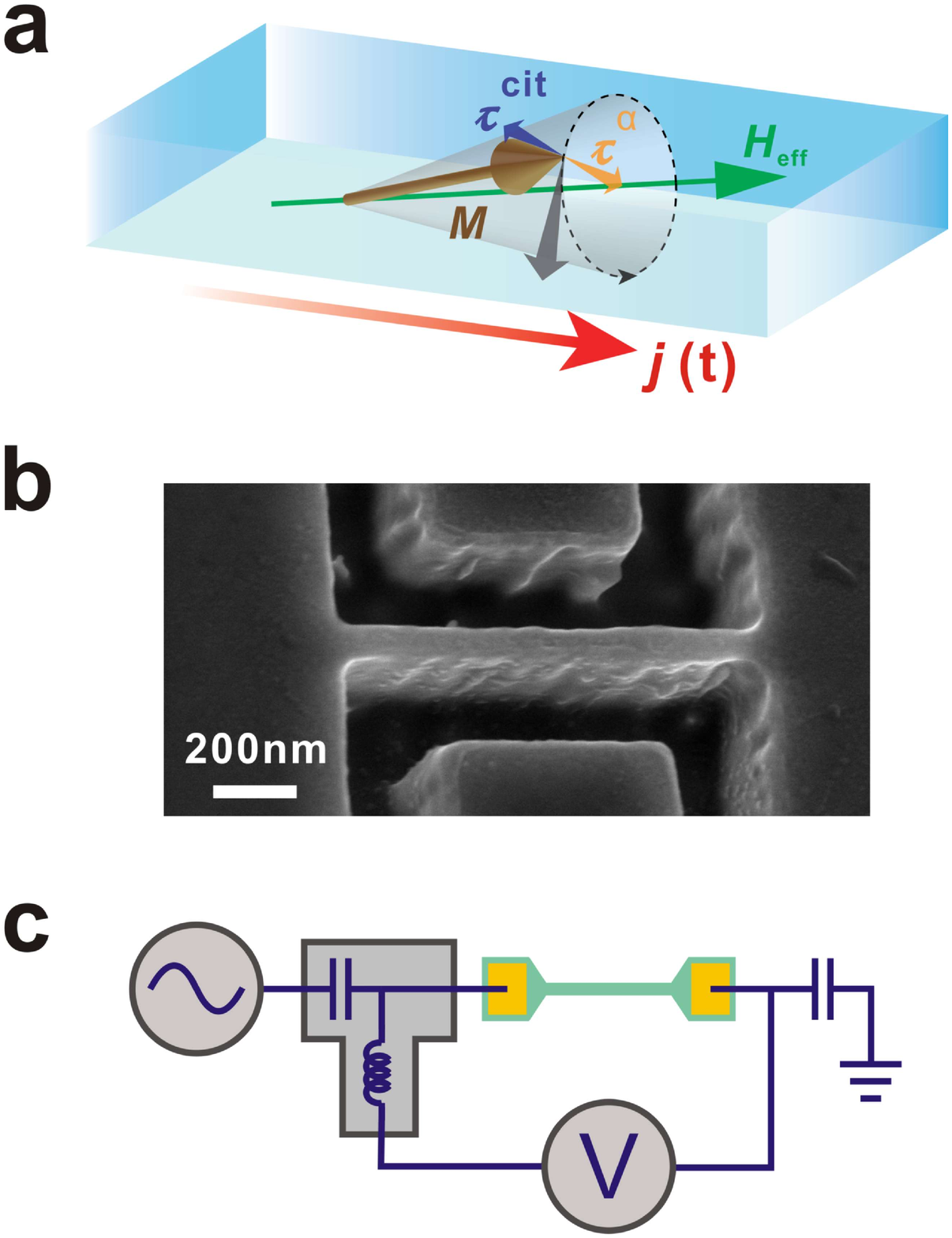}
\end{figure}

\textbf{Figure 1, Principle of the experiment and its setup.} \textbf{a,} Precession of the magnetisation vector \textbf{M} around the total magnetic field $\mathbf{H}_\text{tot}$. \textbf{M} is subject to a damping torque $\tau^{\alpha}$ due to energy dissipation, which causes the magnetic motion to relax towards $\mathbf{H}_\text{tot}$. The driving torque $\tau^\text{SO}$ due to current-induced effective field counters the effect of damping, and leads to steady-state motion $\partial\mathbf{M}/\partial t = -\gamma \mathbf{M} \times \mathbf{H}_\text{tot}$. \textbf{b,} SEM image of a 80~nm-wide bar, patterned from the (Ga,Mn)(As,P) wafer. \textbf{c,} Schematic of the experimental setup.
\newpage

\begin{figure}
	\centering
		\includegraphics[width=15cm]{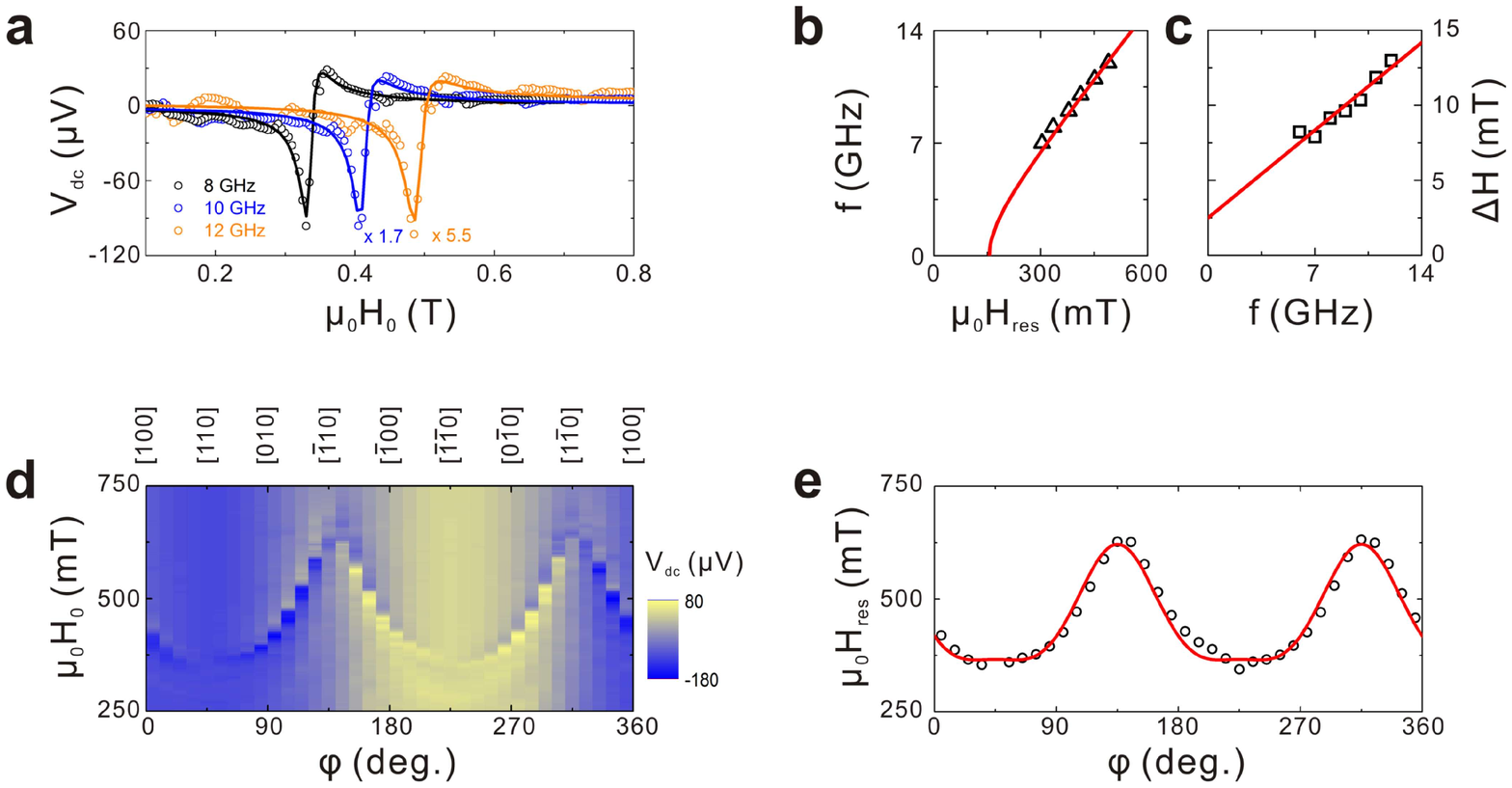}
\end{figure}

\textbf{Figure 2, Spin-orbit driven ferromagnetic resonance.} \textbf{a,} $V_\text{dc}$ measured at 8, 10 and 12~GHz (circles) on the 80~nm-wide device. The resonance peaks are clearly observed and can be well-described by Equation~\ref{eq:Vdc} (solid lines are the fitted results). The difference in the signal level at different $\omega$ is caused by the frequency-dependent attenuation of the microwave circuit. \textbf{b,} The resonance field $H_\text{res}$ as a function of the microwave frequency (black triangles). The red solid line is the fitted results to Equation~\ref{eq:resonance}. \textbf{c,} Frequency-dependence of the FMR linewidth $\Delta H$ (black squares). The data are fitted to a straight line to extract information on $\Delta H_\text{inhomo}$ and $\alpha$. \textbf{d,} $V_\text{dc}$ measured from in-plane rotational scans of the external field $\mathbf{H}_0$. The colour scale represents the magnitude of the voltage. $\varphi$ is the angle between the magnetisation vector $\mathbf{M}$ and the [100] crystalline axis. \textbf{e,} Angle-plot of the resonance field $H_\text{res}$, which is extracted by fitting to each FMR peak using Equation~\ref{eq:Vdc} (black circles). The red line is a fitting curve to Equation~\ref{eq:resonance} to calculate the magnetic anisotropy.

\newpage
\begin{figure}
	\centering
		\includegraphics[width=15cm]{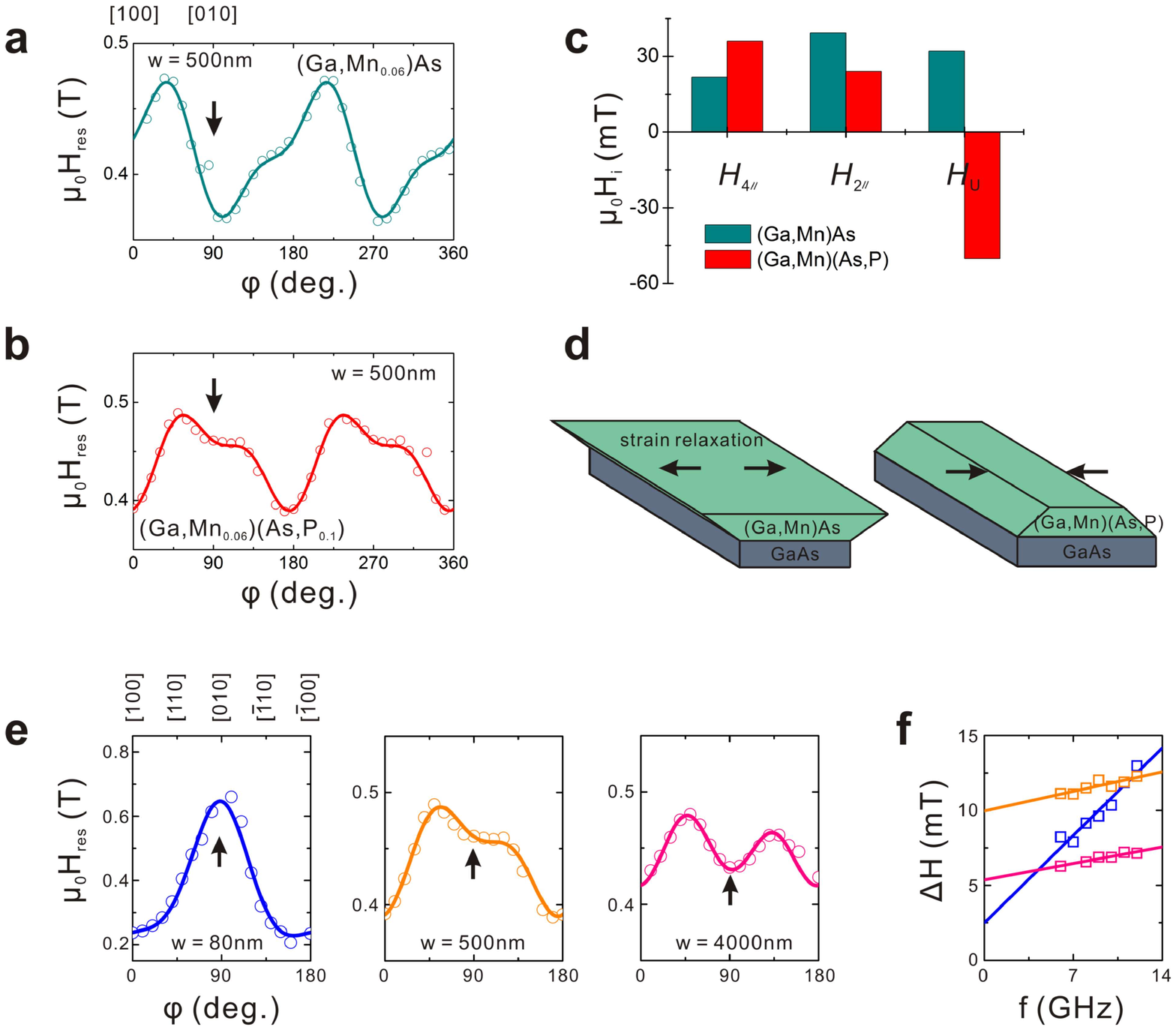}
\end{figure}

\textbf{Figure 3, SO-FMR on devices patterned from different materials and with various sizes.} \textbf{a,} $H_\text{res} (\varphi)$ measured from an in-plane rotational scan on a 500~nm-wide (Ga,Mn$_{0.06}$)As bar (patterned along the [010] axis). The circles are measurement data, and the solid line is the fitted results to Equation~\ref{eq:resonance}. The black arrow marks the long axis of the nano-bar. \textbf{b,} $H_\text{res} (\varphi)$ measured on a (Ga,Mn$_{0.06}$)(As,P$_{0.1}$) device with identical shape and orientation. \textbf{c,} Comparison of the in-plane anisotropy fields $H_i$ between the two samples. \textbf{d,} Schematic of the strain relaxation in the compressively-strained (Ga,Mn)As and and tensile-strained (Ga,Mn)(As,P) nanostructures. \textbf{e,} Comparison of the magnetic anisotropy (in terms of the profiles of $H_\text{res}$) among 80, 500 and 4000~nm-wide (Ga,Mn)(As,P) bars. \textbf{f,} The linewidth $\Delta H$  of the FMR signals measured on the three devices.

\newpage
\begin{figure}
	\centering
		\includegraphics[width=15cm]{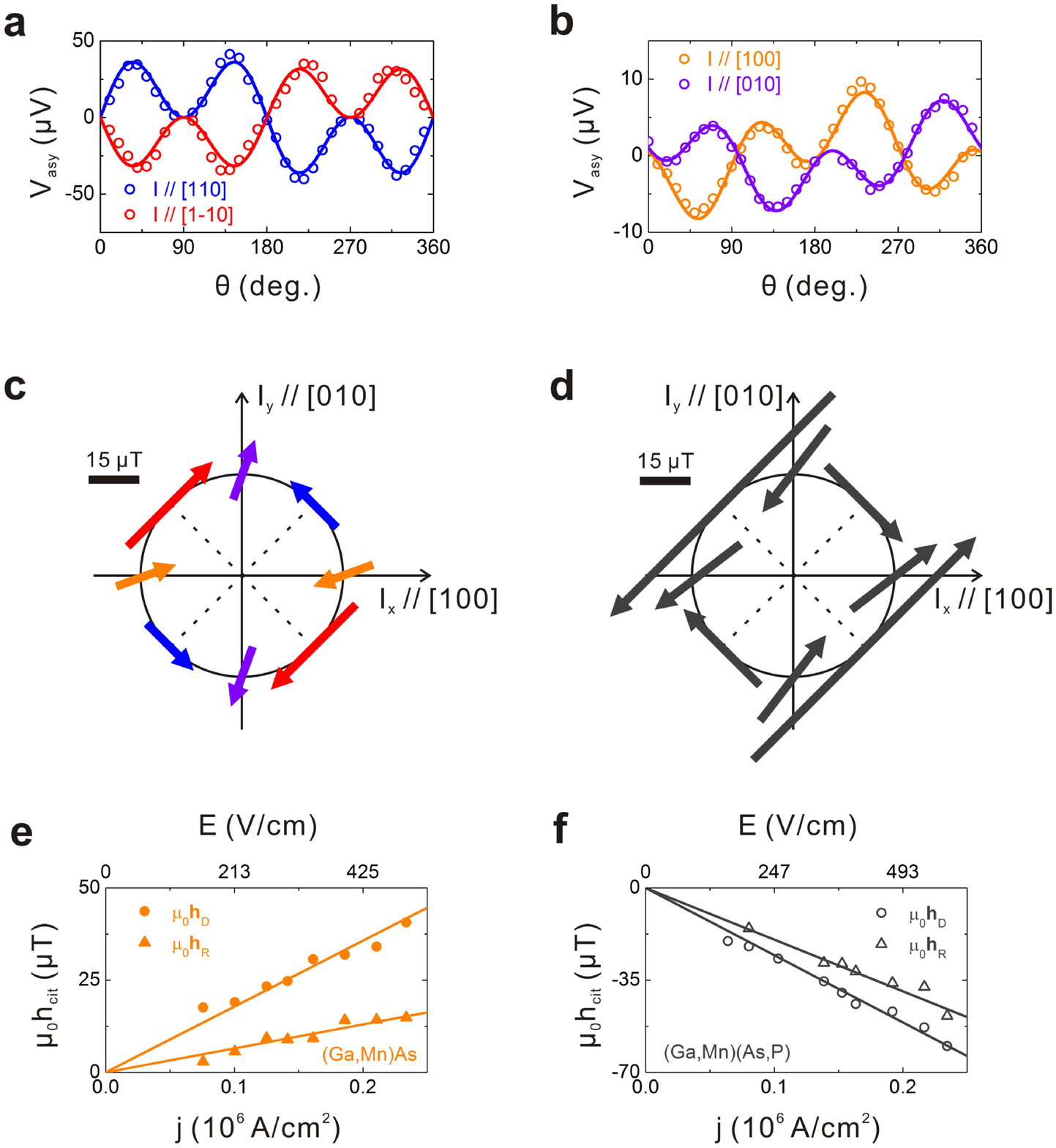}
\end{figure}

\textbf{Figure 4, Characterisation of the driving field in both (Ga,Mn)As and (Ga,Mn)(As,P) devices.} \textbf{a--b,} Amplitudes of the anti-symmetric part of the FMR signal $V_\text{asy}$, measured on a group of 500~nm-wide (Ga,Mn)As bars (circles), patterned along different crystalline directions. The solid lines are fitted results to Equation~\ref{eq:Vasy}. \textbf{c,} Plot of the magnitude and direction of the current-induced effective field $\mathbf{h}_\text{eff}$ measured on the (Ga,Mn)As nano-bars, scaled for a current density $j = 10^5$~A/cm$^{2}$. \textbf{d,} Similar plot for $\mathbf{h}_\text{eff}$ measured on the (Ga,Mn)(As,P) devices. \textbf{e--f,} Current density dependence of $\mathbf{h}_\text{D}$ and $\mathbf{h}_\text{R}$ in both (Ga,Mn)As and (Ga,Mn)(As,P) nano-bars. A second horizontal scale is included for the electric field, calculated from the device resistance (values given in Methods).

\end{document}